\documentclass[prl,twocolumn,showpacs]{revtex4}
\usepackage{amsmath}
\usepackage{dcolumn}
\usepackage{graphicx}

\begin{document}

\title{Universal linear-temperature dependence of static magnetic
susceptibility in iron-pnictides}
\author{G. M. Zhang$^{1}$}
\email{gmzhang@tsinghua.edu.cn}
\author{Y. H. Su$^{2}$}
\author{Z. Y. Lu$^{3}$}
\author{Z. Y. Weng$^{4}$}
\author{D. H. Lee$^{5}$}
\author{T. Xiang$^{6,7}$}
\email{txiang@aphy.iphy.ac.cn}
\affiliation{$^{1}$Department of Physics, Tsinghua University, Beijing 100084, China}
\affiliation{$^{2}$Department of Physics, Yantai University, Yantai 264005, China}
\affiliation{$^{3}$Department of Physics, Renmin University of China, Beijing 100872,
China}
\affiliation{$^{4}$Center for Advanced Study, Tsinghua University, Beijing 100084, China}
\affiliation{$^{5}$Department of Physics, University of California at Berkeley, Berkeley,
CA 94720, USA}
\affiliation{$^{6}$Institute of Physics, Chinese Academy of Sciences, Beijing 100190,
China}
\affiliation{$^{7}$Institute of Theoretical Physics, Chinese Academy of Sciences, Beijing
100190, China}
\date{\today }

\begin{abstract}
A universal linear-temperature dependence of the uniform magnetic
susceptibility has been observed in the nonmagnetic normal state of
iron-pnictides. This non-Pauli and non-Curie-Weiss-like paramagnetic
behavior cannot be understood within a pure itinerant picture. We argue that
it results from the existence of a wide antiferromagnetic fluctuation window
in which the local spin-density-wave correlations exist but the global
directional order has not been established yet.
\end{abstract}

\maketitle

The recent discovery\cite{kamihara} of superconductivity in LaFeAsO$_{1-x}$F$%
_{x}$ has generated strong interest on the investigation of iron-based
pnictide materials. There are mainly two types of materials synthesized: the
rare-earth pnictide oxide layered systems, ReFeAsO denoted as ''1111'' and
the so-called ''122'' systems, MFe$_{2}$As$_{2}$ with M=Ca, Ba, Sr, etc.
Both the 1111 and 122 parent compounds are metals and have shown a spin
density wave (SDW) ordering at $T\sim 130$K, accompanying a
tetragonal-orthorhombic structure phase transition\cite{cruz}. The fact that
the parent compounds of the iron pnictides are antiferromagnetic (AF) has
attracted lots of attention, because of the close analogy with the cuprates.
Indeed, in the most interesting scenario, this suggests that the AF
correlation is intimately connected to the high $T_{c}$ in both materials.
Therefore, a deeper understanding of the AF correlation in the iron
pnictides is of particular importance. The purpose of this paper is to take
a first step in this direction.

In order to establish a microscopic theory for these materials, two
different scenarios, starting from either the weak or strong coupling limit,
have been proposed. The first one invokes an itinerant electron approach in
which the commensurate SDW ordering as well as the structural transition is
believed to be solely induced by the Fermi surface nesting\cite%
{fang-dai,mazin}. In contrast, the second one emphasizes an As-bridged
superexchange antiferromagnetic interactions between the nearest and next
nearest neighboring local moments of irons, which serve as the basic driving
force for both transitions\cite{zhong-yi} without the critical involvement
of the Fermi surface nesting. To distinguish the above two scenarios,
understanding of the origin of the SDW ordering is the key.

Like any ordering phenomena, one can use an order parameter $\vec{n}$ to
describe the SDW order of iron pnictides. In the simplest mean-field
picture, $\vec{n}$ is independent of space and time. Above $T_{\text{SDW}}$,
$|\vec{n}|=0$ and there is no trace of magnetism whatsoever. At $T_{\text{SDW%
}}$ two things occur simultaneously: a finite $|\vec{n}|$ develops and the
directional long range order establishes. In a more realistic picture $\vec{n%
}$ is space (and time) dependent. Above $T_{\text{SDW}}$ even though locally
$|\vec{n}|>0$, due to the lack of directional order, global
antiferromagnetism is absent. In the latter picture the SDW transition is
controlled by the onset of directional long range order. In the following we
shall refer to this as ''SDW moment fluctuation scenario''.

When applying the mean-field picture to the iron pnictides, one expects
normal metallic behavior with no trace of antiferromagnetic correlation
above $T_{\text{SDW}}$. As a result the uniform magnetic susceptibility, $%
\chi _{u}$, should be Pauli paramagnetic like. The $\chi _{u}$ for both the
1111 and 122 compounds are shown in Fig.1 as a function of temperature \cite%
{xhchen2,xhchen3,canfield,klingele,wang-luo}. Interestingly they exhibit a
universal linear temperature dependence in both the undoped and the F-doped
LaFeAsO$_{1-x}$F$_{x}$ \cite{klingele} compounds. Clearly, this is
inconsistent with the mean-field approach expectation.

\begin{figure}[tbp]
\includegraphics [angle=90,width=0.95\columnwidth,clip]{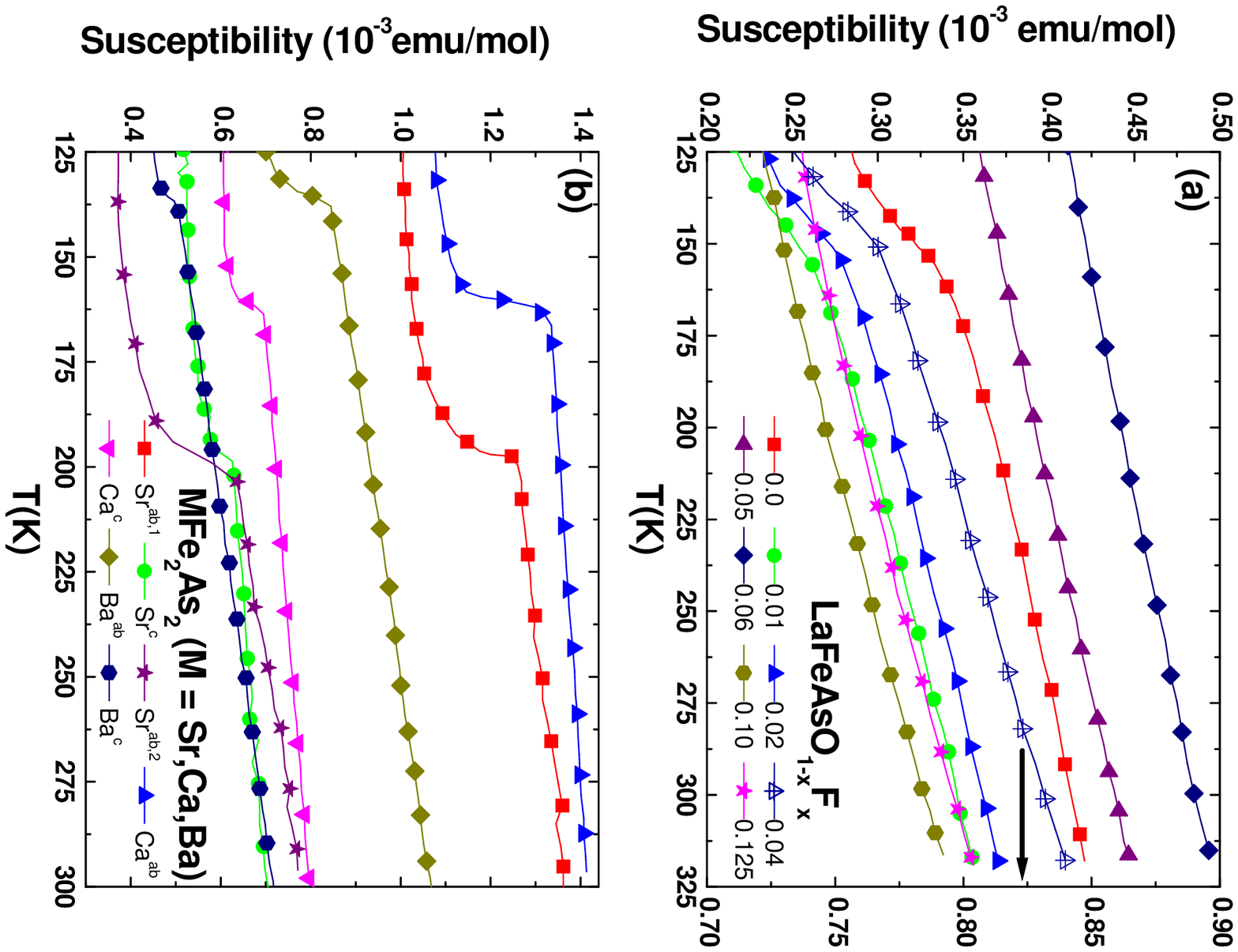}
\caption{Static magnetic susceptibility $\protect\chi _{u}$ vs. temperature.
The experimental data for LaFeAsO$_{1-x}$F$_{x}$ are quoted from Ref.%
\protect\cite{klingele}, SrFe$_{2}$As$_{2}$ from Ref.\protect\cite{canfield}
as 1 and Ref.\protect\cite{wang-luo} as 2, CaFe$_{2}$As$_{2}$ from Ref.%
\protect\cite{xhchen2}, and BaFe$_{2}$As$_{2}$ from Ref.\protect\cite%
{xhchen3}. The arrow indicates the experimental date of LaFeAsO$_{1-x}$F$%
_{x} $ with $x=0.04$ sample in terms of the right scale.}
\end{figure}

In the following, we argue that this linear-T susceptibility is a strong
evidence for the existence of a wide antiferromagnetic fluctuation window of
local magnetic moments. It is important to emphasize, however, that the
metallic behavior of these compounds makes the present local magnetic
moments not \textit{quantized} as those local atomic moment as in a Mott
insulator. Interestingly, in undoped or highly underdoped cuprates La$_{2-x}$%
Sr$_{x}$CuO$_{4}$, $\chi _{u}$ increases linearly with temperature before
reaching a broad peak at a temperature $T_{\max }$ \cite{nakano} just like
iron pnictides. Moreover, the experimental curves can be scaled onto a
universal curve independent of doping. This universal curve agrees with the
theoretical result \cite{sachdev,makivic,troyer} obtained for the
two-dimensional Heisenberg model with nearest neighbor AF coupling.

It is important to note that there are metallic SDW systems which also show
the linear-T susceptibility above $T_{\text{SDW}}$. The best example is
chromium and some of its alloys\cite{Fawcett-1994}. In the case of pure Cr,
diffusive commensurate AF magnetic scattering peak had been observed up to
temperatures $T>2T_{SDW}$, from which a very small effective magnetic moment
($\mu =0.16\sim 0.28\mu _{B}$) can be exacted\cite{grier}. This suggests
that the local AF SDW correlations extend to rather high temperatures.
Another metallic AF system that shows the above linear-T susceptibility
above $T_{\text{N}}$ is Na$_{0.5}$CoO$_{2}$ which is a poor metal with a N%
\.{e}el transition at $86$K\cite{foo}. Thus the linear-T susceptibility
clearly can not be used as evidence for quantized atomic moment as in Mott
insulators.

Put it simply, such a phenomenon just implies a non mean-field transition
into the SDW ordered state. The temperature range showing linear-T
susceptibility is the fluctuation window in the Ginzburg sense. A more
appropriate way of thinking is through Ginzburg-Landau-Wilson theory which
captures the fluctuation of the SDW order parameter. To mimic such a theory
one can write down an \textit{effective} lattice model of fixed magnitude
spin moments and do statistical mechanics on it. If one takes a classical
antiferromagnetic Heisenberg model on a non-frustrated two dimensional
lattice, it can be shown that the above linear-T susceptibility exists in
the temperature range $0<T<T_{\text{MF}}$ with $k_{B}T_{\text{MF}}$ of order
the nearest neighbor exchange constant\cite{Garanin-2000}.

In the present paper, we prefer to start from a quantum spin model and do
finite temperature statistical mechanics. The following two dimensional
frustrated antiferromagnetic $J_{1}-J_{2}$ Heisenberg model is assumed
\begin{equation}
H=J_{1}\sum_{\langle i,j\rangle }S_{i}\cdot S_{j}+J_{2}\sum_{\langle \langle
i,j\rangle \rangle }S_{i}\cdot S_{j},
\end{equation}%
where $\langle i,j\rangle $ and $\langle \langle i,j\rangle \rangle $ denote
the summations over the nearest and next nearest neighbors, respectively.
With $J_{2}>J_{1}/2$, this model captures the $(\pi ,0)$ and $(0,\pi )$
ordering tendencies of iron-pnictides. Here we assume that at $T=0$K the
spins are ferromagnetic ordering along the x-direction and antiferromagnetic
ordering along the y-direction. So the lattice is bipartite and divided into
A and B sublattices. On the A (B) sublattice, the vacuum state is the $%
S^{z}=S$ ($-S$) state. There are two spins in each unit cell.

We then use the antiferromagnetic Dyson-Maleev transformation to represent
the spin operators. 
Different from the variational approach used by Takahashi \cite{takahashi},
we approximate the model Hamiltonian by keeping the \textit{quadratic}
interactions of the boson operators only. Then the model Hamiltonian is
hermitian, and it can be expressed after Fourier transform as \cite{yue-hua}
\begin{eqnarray}
H &\approx &\sum_{k}\left[ \eta _{k}\left( a_{k}^{\dagger
}a_{k}+b_{k}^{\dagger }b_{k}\right) +\Lambda _{k}\left(
a_{k}b_{-k}+a_{k}^{\dagger }b_{-k}^{\dagger }\right) \right]  \notag \\
&&-2NS(J_{2}S+\lambda )
\end{eqnarray}%
where $\eta _{k}=2J_{1}S\cos k_{x}+2J_{2}S+\lambda $, $\Lambda
_{k}=2J_{1}S\cos \frac{k_{y}}{2}+2J_{2}S\gamma _{k}$, $\gamma _{k}=\cos
k_{x}\cos \frac{k_{y}}{2}$, and a chemical potential term $\lambda $ has
been introduced to make the local magnetization vanish at finite
temperatures. By using the Bogoliubov transformation, we can diagonalize the
Hamiltonian as%
\begin{equation}
H=\sum_{k}\epsilon _{k}\left( \alpha _{k}^{\dagger }\alpha _{k}+\beta
_{-k}^{\dagger }\beta _{-k}\right) +\varepsilon _{g}N,
\end{equation}%
where $\epsilon _{k}=\sqrt{\eta _{k}^{2}-\Lambda _{k}^{2}}$ and $\varepsilon
_{g}$ is the ground state energy per site. From the free energy, the
chemical potential $\lambda $\ is determined by the following equation
\begin{equation}
\frac{1}{N}\sum_{k}\frac{\eta _{k}}{\epsilon _{k}}\coth \left( \frac{%
\epsilon _{k}}{2k_{B}T}\right) =2S+1.
\end{equation}%
Moreover, when a magnetic field is applied, the above treatment can still be
carried out by including the Zeeman terms, and the static uniform magnetic
susceptibility is derived as
\begin{equation}
\chi _{u}=\frac{\left( gu_{B}\right) ^{2}}{4k_{B}TN}\sum_{k}\frac{1}{\sinh
^{2}\left( \frac{\epsilon _{k}}{2k_{B}T}\right) }.
\end{equation}%
However, compared to the static uniform susceptibility expression obtained
from the correlation function, a factor of $\frac{1}{3}$ has to be
multiplied.\cite{yue-hua}

Numerical calculations for the static uniform magnetic susceptibility $\chi
_{u}$ can be performed at finite temperatures. Surprisingly, it has been
found that $\chi _{u}$ behaves as linearly temperature dependence before it
reaches a broad peak, then it can be fit as the Curie-Weiss behavior.\cite%
{yue-hua} There is a clear crossover regime connecting these two different
regimes. Moreover, as the coupling ratio of $J_{2}/J_{1}$ is increased, the
window of the linear magnetic susceptibility becomes wider. In other words,
the maximal value of the broad peak is also shifted as increasing the
coupling ratio of $J_{2}/J_{1}$. Of course, such a treatment is just a
qualitative description of the nonmagnetic state of this frustrated
Heisenberg model. In Fig.2, we present the numerical results of $\chi _{u}$
at finite temperatures for $S=1$ and $J_{2}/J_{1}=$ $1.0$, $1.5$. For the
case of $J_{2}/J_{1}=1.0$, the uniform susceptibility in the temperature
range between $0\sim 0.9J_{1}$, $\chi _{u}$ can be fit as
\begin{equation}
\chi _{u}=\chi _{0}\left[ 1+a\left( \frac{T}{J_{1}}\right) \right] ,\text{ \
\ \ \ }a>0.
\end{equation}%
Quantitatively by taking $J_{1}\sim J_{2}=55\mathrm{\ meV}$
estimated by the local density approximation (LDA)
calculation\cite{zhong-yi}, we find $\chi _{0}\sim 3\times
10^{-4}$ emu$/$mole, which is very close to the experimental
values extrapolated from the linear-T regime in Fig. 1.
\begin{figure}[tbp]
\includegraphics [angle=90,width=0.95\columnwidth,clip]{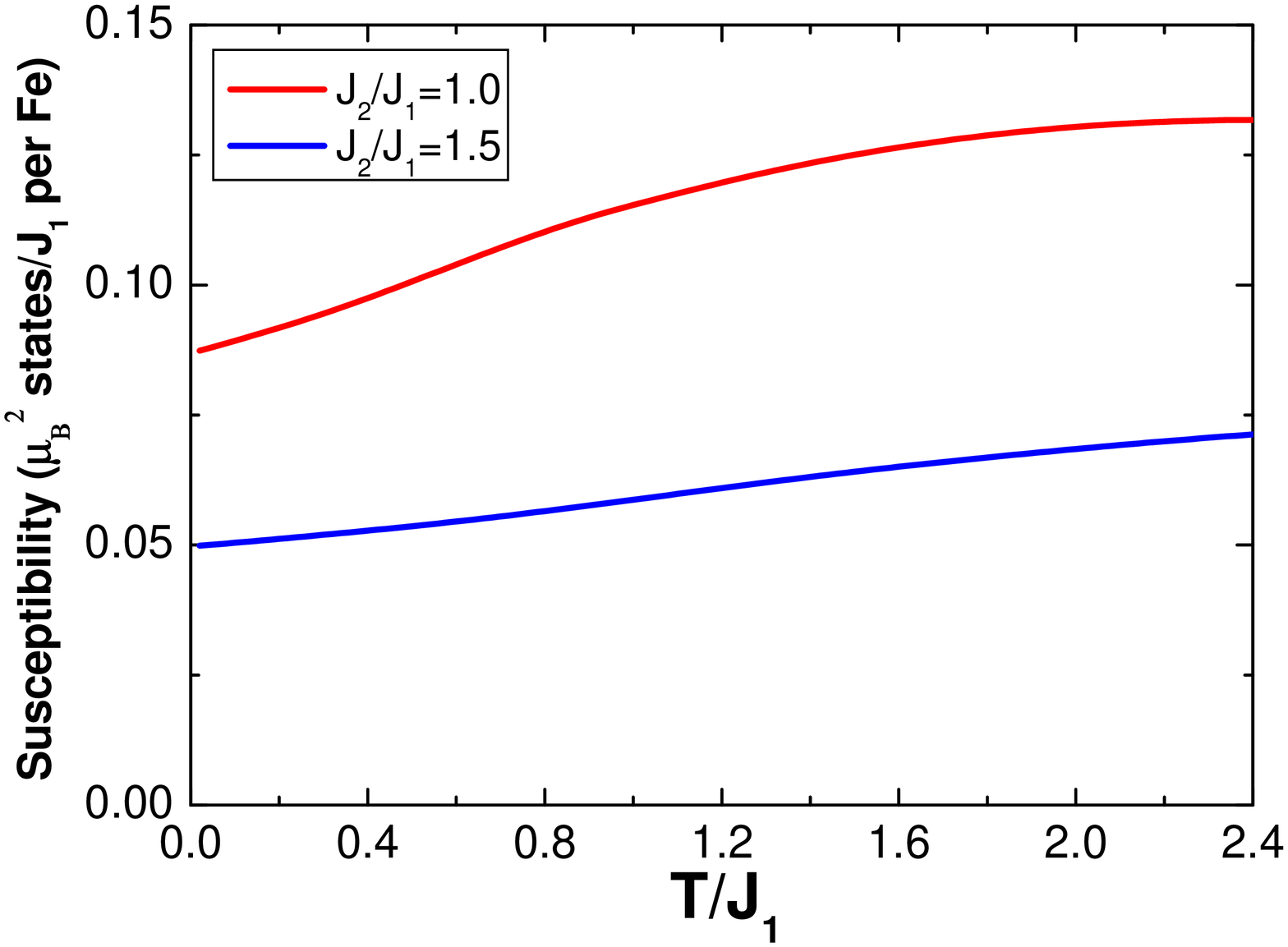}
\caption{Static uniform magnetic spin susceptibility $\protect\chi $ as a
function of temperature obtained from the Dyson-Maleev linear spin-wave
calculation for $S=1$ case.}
\end{figure}

It is noted that a wide range of temperatures showing linear-T
susceptibility can be attributed to the Mermin-Wagner theorem, which says
that a two-dimensional Heisenberg system can not order at non-zero
temperature. As a result, all temperatures below the mean-field crossover
are in the fluctuation regime. The Curie-Weiss-like behavior will eventually
recover at higher temperatures beyond $T_{\max }\sim J_{2}/k_{\mathrm{B}}$,
where the correlation length is less than a lattice constant and the moments
become effectively free, similar to the cuprates\cite{nakano,Gu-Weng}. Of
course the true iron-pnictide systems show a finite temperature SDW ordering
transition. This is due to the inter-layer coupling $J_{z}$. In this case,
we expect the fluctuation window to lie between the mean-field crossover and
the $T_{\text{SDW}}$. In addition, due to the presence of competing
interactions ($J_{1}$ and $J_{2}$), it is argued that above the SDW ordering
transition, there should be an Ising-like transition where the symmetry
between the ($\pi ,0$) and ($0,\pi $) SDW patterns are broken. Such an
transition necessarily breaks the lattice rotation symmetry, and as a result
can trigger the tetragonal-orthorhombic distortion\cite{jphu,cenker-sachdev}%
. The persistence of the linear-T susceptibility into the doped regime
implies that the SDW correlation is strong in the superconducting samples.
This can be used as indirect evidence for the involvement of
antiferromagnetic correlation in Cooper pairing.

We can not overemphasize that the above quantum model is merely used to
mimic the Ginzburg-Landau-Wilson description of the SDW moment fluctuation
scenario. It should not be used to implies that we believe quantized $S=1$
atomic moment exists in the system.

Apparently, there is coupling between the SDW moments and conduction
electrons near the Fermi surface. For example, the SDW transition induces
abrupt changes of the Drude weight\cite{nanlin}, magneto-resistance\cite%
{haihu}, and Hall coefficient\cite{haihu}. In addition, angle-resolved
photoemission experiment has shown a change of the electronic structure near
the Fermi energy at $T_{\text{SDW}}$\cite{feng}. These experiments suggest
that as the SDW moment orders, magnetic scattering further gap out parts of
the Fermi surface and as a result some itinerant carriers are lost.

Let us now switch to the electron origin of the SDW order. A popular point
on this issue says that the SDW moments form because of the Fermi surface
nesting effect\cite{fang-dai,mazin}. First of all, the Fermi surface nesting
is an ``instability'' concept. To be precise, in the presence of Fermi
surface nesting even \textit{infinitely weak} SDW channel quasiparticle
scattering can open the SDW gap. For a strong scattering, however, nesting
is not required. According to the band calculations, the Fermi surface are
not well nested by the magnetic ordering wave vectors $(\pi ,0)$ or $(0,\pi
) $. In addition, the long-ranged ordering moment, which is a lower bound of
the preformed SDW moment, is about $0.365\mu _{B}$ for LaFeAsO \cite{cruz},
and $0.873\mu _{B}$ for BaFe$_{2}$As$_{2}$ \cite{huang}. These moments are
rather big since they are comparable with the $T=0$ ordering moment $0.6\mu
_{B}$ of the spin-1/2 Heisenberg model on the square lattice.

Such a large magnetization moment also rules out the SDW transition being
the mean-field SDW moment formation temperature. If that were the case, one
expects only the electronic states at energy $~k_{B}T_{\text{SDW}}$ away
from the Fermi energy would be perturbed. Given $T_{\text{SDW}}\sim 130$K
and the band structure results, we estimate an upper bound of the ordering
moment to be $\sim 0.02\mu _{B}$, which is more than one order of magnitude
less than the measured value.

In our opinion, the SDW fluctuation moment is more likely due to the strong
short-range repulsion between the electrons. For example, Ref.\cite{zhong-yi}
emphasizes an As-bridged antiferromagnetic interactions between the nearest
and next nearest neighbor iron electrons, which serve as the basic driving
force for SDW moment formation without the critical involvement of the Fermi
surface nesting. In addition, Ref.\cite{Ying} takes the un-nested LDA band
structure adding moderate strong Hubbard-like and Hund-like interactions,
and obtains a good fraction of $\mu _{B}$ for the SDW ordering moment in an
mean-field theory\cite{Ying}. Finally the LDA-based SDW mean-field
calculations have yielded the ordering moment between $2.2\mu _{B}$ and $%
2.6\mu _{B}$\cite{zhong-yi}. However it is typical that all such mean-field
calculations overestimate the ordering moment since it does not capture the
long-wave-length directional fluctuations.

The experimental evidences as well as the theoretical considerations all
lead us to conclude that the SDW moment formation temperature for the
iron-pnictides materials should occur at much higher temperature than $T_{%
\text{SDW}}$. Thus there should be a ''psudogap'' temperature for iron
pnictides as well. Below such a psudogap temperature, it is appropriate to
consider an effective lattice spin model with fixed moments such as the one
given by Eq.(1) to describe the magnetic properties of the system. By
comparing the energy of a variety of magnetic structures, Ma, \textit{et.
al. }\cite{zhong-yi} have estimated $J_{1}\sim J_{2}$ to be about $55$meV
for LaFeAsO and $35$meV for BaFe$_{2}$As$_{2}$.

In conclusion, we have argued that the universal linear temperature
dependence of the susceptibility provides a strong evidence for the SDW
fluctuation moments with strong antiferromagnetic interactions above the SDW
transition temperature in iron-pnictides. This linear susceptibility can be
effectively described as the finite temperature behavior of a Heisenberg
model with nearest and next nearest-neighbor AF interactions. Further
investigations are certainly needed to put our conclusion on a solid ground.

\begin{acknowledgments}
This work is partially supported by NSFC-China and the National Program for
Basic Research of MOST, China.
\end{acknowledgments}


\begin{thebibliography}{99}
\bibitem{kamihara} Y. Kamihara, T. Watanabe, M. Hirano, and H. Hosono, J.
Am. Chem. Soc. \textbf{130}, 3296 (2008).

\bibitem{cruz} C. de la Cruz, Q. Huang, J. W. Lynn, J. Li, W. Ratcliff II,
J. L. Zarestky, H. A. Mook, G.F. Chen, J. L. Luo, N. L. Wang, and P. Dai,
Nature (London) \textbf{453}, 899 (2008).

\bibitem{fang-dai} J. Dong, H. J. Zhang, G. Xu, Z. Li, G. Li, W. Z. Hu, D.
Wu, G. F. Chen, X. Dai, J. L. Luo, Z. Fang, and N. L. Wang, Europhys. Lett.
\textbf{83}, 27006 (2008).

\bibitem{mazin} I. I. Mazin, D. J. Singh, M. D. Johannes, and M. H. Du,
Phys. Rev. Lett. \textbf{101}, 057003 (2008).

\bibitem{zhong-yi} F. Ma, Z.Y. Lu, and T. Xiang, Phys. Rew. B \textbf{78},
224517 (2008); arXiv:0806.3526.


\bibitem{xhchen2} G. Wu, H. Chen, T. Wu, Y. L. Xie, Y. J. Yan, R. H. Liu, X.
F. Wang, J. J. Ying, and X. H. Chen, J. Phys. Condensed Matter \textbf{20},
422201 (2008).

\bibitem{xhchen3} X. F. Wang, T. Wu, G. Wu, H. Chen, Y. L. Xie, J. J. Ying,
Y. J. Yan, R. H. Liu, and X. H. Chen, arXiv:0806.2452.

\bibitem{canfield} Y. Q. Yan, A. Kreyssig, S. Nandi, N. Ni, S. L. Bud'ko, A.
Kracher, R. J. McQueeney, R. W. McCallum, T. A. Lograsso, A. I. Goldman, and
P. C. Canfield, Phys. Rev. B. \textbf{78}, 024516 (2008).

\bibitem{klingele} R. Klingeler, N. Leps, I. Hellmann, A. Popa, C. Hess, A.
Kondrat, J. Hamann-Borrero, G. Behr, V. Kataev, and B. Buechner,
arXiv:0808.0708.

\bibitem{wang-luo} G. F. Chen, J. L. Luo and N. L. Wang, unpublished.

\bibitem{nakano} T. Nakano, M. Oda, C. Manabe, N. Momono, Y. Miura, and M.
Ido, Phys. Rev. B \textbf{49}, 16000 (1994).

\bibitem{sachdev} A. V. Chubukov and S. Sachdev, Phys. Rev. Lett. \textbf{71}%
, 169 (1993).

\bibitem{makivic} M. Makivic and H.-Q. Ding, Phys. Rev. B \textbf{43}, 3562
(1991).

\bibitem{troyer} J.-K. Kim and M. Troyer, Phys. Rev. Lett. \textbf{80}, 2705
(1998).

\bibitem{Fawcett-1994} E. Fawcett, Rev. Mod. Phys. \textbf{60}, 209 (1988);
E. Fawcett, H. L. Alberts, V. Yu Galkin, D. R. Noakes, and J. V. Yakhmi,
Rev. Mod. Phys. \textbf{66}, 25 (1994).

\bibitem{grier} B. H. Grier, G. Shirane, and S. A. Werner, Phys. Rev. B
\textbf{31}, 2892 (1985).

\bibitem{foo} M. L. Foo, Y. Wang, S. Watauchi, H. W. Zandbergen, T. He, R.
J. Cava, N. P. Ong, Phys. Rev. Lett., \textbf{92}, 247001 (2004).

\bibitem{Garanin-2000} D. Hinzke, U. Nowak, and D.A. Garanin, Eur. Phys. J.
B \textbf{16}, 435 (2000).


\bibitem{takahashi} M. Takahashi, Phys. Rev. B \textbf{40}, 2494 (1989).

\bibitem{yue-hua} Y. -H. Su, et. al., in preparation.

\bibitem{Gu-Weng} Z. C. Gu and Z. Y. Weng, Phys. Rev. B \textbf{72}, 104520
(2005).

\bibitem{jphu} C. Fang, H. Yao, W.F. Tsai, J.P. Hu, and S. A. Kivelson,
Phys. Rev. B \textbf{77}, 224509 (2008).

\bibitem{cenker-sachdev} C. Xu, M. Muller, and S. Sachdev, Phys. Rev. B
\textbf{78}, 020501 (R) (2008).

\bibitem{nanlin} W. Z. Hu, J. Dong, G. Li, Z. Li, P. Zheng, G. F. Chen, J.
L. Luo, and N. L. Wang, Phys. Rev. Lett. \textbf{101}, 257005 (2008).

\bibitem{haihu} P. Cheng, H. Yang, Y. Jia, L. Fang, X. Zhu, G. Mu, H.-H.
Wen, Phys. Rev. B \textbf{78}, 134508 (2008).

\bibitem{feng} L. X. Yang, H. W. Ou, J. F. Zhao, Y. Zhang, D. W. Shen, B.
Zhou, J. Wei, F. Chen, M. Xu, C. He, X. F. Wang, T. Wu, G. Wu, Y. Chen, X.
H. Chen, Z. D. Wang, and D. L. Feng, arXiv:0806.2627.

\bibitem{huang} Q. Huang, Y. Qiu, W. Bao, J.W. Lynn, M.A. Green, Y.C.
Gasparovic, T. Wu, G. Wu, and X. H. Chen, Phys. Rev. Lett. \textbf{101},
257003 (2008).

\bibitem{Ying} Y. Ran, F. Wang, H. Zhai, A. Vishwanath, D.-H. Lee,
arXiv:0805.3535.












\end{thebibliography}
\end{document}